\documentstyle[11pt,aaspp4,amssym,epsfig]{article}

\def\simg{\mathrel{\hbox{\rlap{\lower.55ex \hbox {$\sim$}}
                   \kern-.3em \raise.4ex \hbox{$>$}}}}
\def\siml{\mathrel{\hbox{\rlap{\lower.55ex \hbox {$\sim$}}
                   \kern-.3em \raise.4ex \hbox{$<$}}}}
\def\Mesz{M\'esz\'aros~}

\def\beq{\begin{equation}}
\def\enq{\end{equation}}
\def\bea{\begin{eqnarray}}
\def\ena{\end{eqnarray}}
\def\bec{\begin{center}}
\def\enc{\end{center}}

\def\L52{L_{52}}
\def\E53{E_{53}}
\def\ro{r_o}

\def\rpi{r_\pi}

\def\etapi{\eta_\pi}
\def\etapish{\eta_{\pi ,sh}}
\def\etast{\eta_\ast}

\def\st{\sigma_T}

\def\msun{M_\odot}
\def\eps{\epsilon}
\def\varep{\varepsilon}

\def\nue{\nu_e}
\def\num{\nu_\mu}
\def\barnu{\bar\nu}
\def\barnue{\bar \nu_e}
\def\barnum{\bar \nu_\mu}

\begin{document}

{\it \hfill ApJL, subm. 7/6/2000} 

\title{ Multi-GeV Neutrinos from Internal Dissipation \\
         in GRB Fireballs } 


\author{P. \Mesz$^{1,2}$ \& M.J. Rees$^1$ }
\smallskip\noindent
$^1${Institute of Astronomy, University of Cambridge, Madingley Road, Cambridge
CB3 0HA, U.K.}\\
\smallskip\noindent
$^2${Dpt. of Astronomy \& Astrophysics, Pennsylvania State University,
University Park, PA 16803}

\begin{abstract}

Sub-photospheric internal shocks and transverse differences of the bulk Lorentz factor
in relativistic fireball models of GRB lead to neutron diffusion relative to protons, 
resulting in inelastic nuclear collisions. This results in significant fluxes of 
$\num(\barnum)$ of $\sim 3$ GeV and $\nue(\barnue)$ of $\sim 2$ GeV, scaling with the
flow Lorentz factor $\eta < \etapi \sim 400$. This extends significantly the parameter space 
for which neutrinos from inelastic collision are expected, which in the absence of the above 
effects requires values in excess of $\etapi$.  A model with sideways diffusion of neutrons 
from a slower wind into a fast jet can lead to production of $\num(\barnum)$ and 
$\nue(\barnue)$ in the 2-25 GeV or higher range, depending on the value of $\eta$.
The emission from either of these mechanisms from GRB at redshifts 
$z\sim 1$ may be detectable in suitably densely spaced detectors. 

\end{abstract}

\keywords{Gamma-rays: Bursts -  Neutrinos - Cosmology: Miscellaneous}

\section{Introduction}

Neutrinos in three different energy ranges are expected from gamma-ray burst sources. 
The most straightforward and also the hardest to detect are quasi-thermal 10-30 MeV neutrinos 
associated with the stellar collapse or merger event that triggers the burst (Kumar 1999, 
Ruffert \& Jahnka 1999). The low energies imply an extremely low  detection cross section, which 
for sources at cosmological distances makes the required detector volume prohibitive. At much
higher energies, $\eps_\nu\simg 2\times 10^{14}$ eV neutrinos are expected from the photopion 
process in $p\gamma$ interactions between protons accelerated in shocks and the GRB photons 
produced by the same shocks associated with the fireball.  These have a much higher 
cross section, and significant detection rates are predicted (Waxman \& Bahcall 1997, Vietri 1998,
Rachen \& \Mesz 1998).
Such neutrino events would directly test the shock acceleration mechanism, since they
require relativistic protons with energies $\sim 10^{16}$ eV. A third type of neutrino
emission, which is expected in the 5-15 GeV energy range (Derishev et al, 1999, Bahcall \& 
\Mesz 2000), occurs only if neutrons are present in the fireball, and the fluences are
detectable in km$^3$ detectors. These neutrinos arise from longitudinal decoupling of the 
$n$ and $p$ flows, and this is expected to occur whether or not shocks occur, provided 
a substantial fraction of neutrons is present in the fireball and the total fireball bulk 
Lorentz factor $\eta \simg \etapi \simeq 400$. 

In this paper we point out that if internal shocks occur below the radiation photosphere, 
this can lead to a regime where rapid diffusion of neutrons both parallel and transverse 
to the radial direction occurs. This leads to neutrinos with energy $\eps_{\nu\barnu} \sim 3
(\eta/\etapish)(2/1+z)$ GeV for a range of $\eta$ both below and above a critical value 
$\etapish \sim \etapi^{4/5}\sim 120$. This extends considerably the parameter space available 
for neutrino production from $np$ interactions, in contrast to the smooth homogeneous fireball 
case where $\eta\simg \etapi$ is required. Whereas in typical shocks responsible for MeV 
gamma-rays the neutrino efficiencies from $pp$ are low due to increased coupling and lower 
optical depths (Paczy\'nski \& Xu 1994), here the neutrino fluence is maximized with $np$ 
collisions in shocks at or below the pionosphere, where the inelastic optical depth is $\simg 1$.
The flux of such neutrinos may be detectable with km$^3$ Cherenkov detectors currently under 
planning, for a sufficiently dense phototube spacing. Similar or higher neutrino energies and 
rates are also possible from bursts involving a slow wind and a fast jet, where neutrons from 
the slow wind diffuse sideways into the highly relativistic GRB beam.

\section{Sub-photospheric shocks and $n$ diffusion}
\label{sec:heating}

In a relativistic GRB outflow consisting of $\gamma,e^\pm,p$ and $n$ which
is initially accelerating with a common bulk Lorentz factor $\Gamma \simeq
r/r_o$, the dynamics is the same for a spherical  geometry of for an outflow
constrained within a channel of angular width $\theta_j \simg \Gamma^{-1}$.
There are two dynamical regimes, depending on whether $\eta=L/{\dot M} c^2$
is greater or less than the critical value 
\beq
\etapi= \left( { L \sigma_\pi \over 4\pi m_p c^3 r_o (1+\xi) }\right)^{1/4}
 \simeq  4.6 \times 10^2 \L52^{1/4} r_{o7}^{-1/4} (1+\xi)^{-1/4}
\label{eq:etapi}
\enq
where $\xi=n_n/n_p$ is the neutron/proton ratio, which for simplicity is taken to be constant 
over the range of interest, $\sigma_\pi\sim 3\times 10^{-26}$ cm$^2$ is the inelastic cross 
section near threshold, and $r_o=10^7 r_{o7}$ cm is the initial radius of the outflow.
The parameter $\etapi$ of equation (\ref{eq:etapi}) can also be written as $\etapi= \eta_\ast 
(\sigma_\pi/\sigma_T)^{1/4}$, where $\st$ is the Thomson cross section and 
\beq
\etast= \left( {L \st \over 4\pi \ro m_p c^3 (1+\xi)} \right)^{1/4} 
 \simeq 10^3 \L52^{1/4}r_{07}^{-1/4}(1+\xi)^{-1/4}~
\label{eq:etast}
\enq
is the critical value of $\eta$ for $\gamma,p$ decoupling in a $p,n$ outflow, which enters 
also in the definition of the photospheric radius (e.g. \Mesz \& Rees 2000 for a pure proton 
fireball). 

For $\eta\geq \eta_\pi$, the radial velocity of the $n$ and $p$ becomes decoupled while both 
components are in the accelerating regime where $\Gamma\propto r$, leading to 
relative drift velocities between $n$ and $p$ of $v_{rel}\to c$, and 5-10 GeV neutrinos 
are expected from inelastic $np$ collisions (Bahcall \& \Mesz, 2000). 
On the other hand, for $\eta <\etapi$ a decoupling of the bulk velocity of the $n$ and $p$ in 
the radial direction is prevented by the fact that the radial motion of both the $n$ and $p$ 
saturate to an asymptotic Lorentz factor $\Gamma_p \sim \Gamma_n \sim \eta$ at $r \simg 
r_s\simeq r_o \eta$ before a relative drift $v_{rel}\to c$ has been reached, and thereafter the 
various components coast together. If $\eta$ is constant in time and uniform across the channel 
of angular width $\theta_j$, 
the comoving temperature $T'=T_o(r_o/r)$ or $T'=(T_o/\eta)(\eta/r)^{2/3}$ for $r/r_o$ below or
above the value $\eta$, and $T_o=1.2 \L52^{1/4} r_{07}^{-1/2}$ MeV is the initial temperature 
at $r_o$ (henceforth denoting with primes quantities measured in the comoving frame).
The thermal velocities of the baryons are therefore subrelativistic and the relative CM 
kinetic energy of the $n$ and $p$ is below the threshold energy $\sim 140$ MeV required for
inelastic nuclear collisions leading to neutrinos. 

However, for $\eta\gtrless \etapi$, it is possible and likely that 
temporal or angular inhomogeneities in $\eta$ do exist in realistic GRB outflows. Both 
types of inhomogeneities imply the presence of density gradients, and thus the possibility 
for neutrons to diffuse into regions whose bulk velocity is significantly different from 
that where they originated. When the neutrons encounter protons or other neutrons with relative 
velocities $v_{rel} \simg 0.6 c$ the CM inelastic threshold energy $\eps_{rel}\simg 140$ MeV 
is reached, leading to pion formation and neutrinos. In order for this to occur with significant 
probability, one requires the inelastic $np$ comoving mean-free path $\ell'_{np}\sim 
(n'_p \sigma_\pi)^{-1}$ to be comparable to the scale of the inhomogeneity in the comoving 
frame, where $n'_p=L/[4\pi r^2 m_p c^3 \Gamma \eta (1+\xi)]$ is the comoving proton density
(henceforth primed quantities are in the comoving frame). In the case of internal shocks for 
which $\Delta\Gamma \sim \Gamma$, which ensures $v_{rel}\to c$, the characteristic length scale 
of the shocks is $r/\Gamma$. In the case of sideways diffusion, a thermal velocity 
$v_{t}\sim v_{rel}\to c$ also requires internal shocks to have re-heated the gas, 
so both effects are tied to each other, and the transverse diffusion is limited to the 
causal angle $\Gamma^{-1}$ so the lengthscale is again $r/\Gamma$. In the saturated regime 
(which is a prerequisite for internal shocks to occur), when the $n,p$ thermal relative 
velocities $v_{rel}\to c$ so $\sigma \to \sigma_\pi$, the inelastic optical depth parallel 
or perpendicular to the shock normal becomes smaller than unity above a radius $r_\pi$ given by
\beq
(r_\pi/r_o) = \etapi (\etapi/\eta)^3~.
\label{eq:rpi}
\enq
In the above expression the $\etapi$ dependence includes a factor $(1+\xi)^{-1/4}$, which 
assumes protons as the only targets, for comparison with the radial decoupling case
(the case where both $n$ and $p$ are targets is given by the same equation (\ref{eq:rpi}) 
by formally setting $\xi=0$ in the $\etapi$ factor).  The internal shocks, required to provide 
$v_{rel}\to c$, occur in a coasting regime, at radii 
\beq
r_{sh}/\ro \sim (t_v/t_o) \eta^2~,
\label{eq:rsh}
\enq
where $t_v$ is the observer-frame variability timescale on which $\eta$ varies significantly, 
and $t_v \simg t_o \sim r_o/c \sim 10^{-3}$ s, where $t_o$ is the minimum dynamical timescale 
in the system over which such variation can occur involving a substantial fraction of the total
outflow energy. This defines a minimum shock radius $r_{sh,m}/r_o \sim \eta^2$, for $\eta 
<\etast$. For radii $r_{sh,m} \siml r \siml r_\pi$ shocks leading to $v_{rel}\to c$ are possible, 
while  the additional condition $\tau'_\pi \geq 1$ ensures efficient pion production and neutrino 
emission. This region is shown by horizontal hashed lines in Figure 1.
The requirement that $r_{sh,m} \siml \rpi$ translates into the condition 
\beq
\eta \leq \eta_{\pi,sh} \sim (t_o/t_v )^{1/5}\etapi^{4/5}\simeq 120 \L52^{1/5}(t_o/t_v)^{1/5}~,
\label{eq:etapish}
\enq
For radii above $r_\pi$, or for values of $\eta \simg \eta_{\pi,sh}$, shocks and inelastic 
collisions can also occur in the saturated regime at $r>r_{sh,m}$, but the mean-free path 
$\ell'_\pi$ is longer than the characteristic system length $r/\Gamma$, and the optical depth 
for inelastic collisions $\tau'_\pi< 1$, being equal to the inverse ratio of these two lengths.
This regime is shown with vertical hashed lines in Figure 1. For $\eta > \etast >\etapi$
the neutrons have decoupled at a terminal value $\Gamma_{nf} \sim (3/4) \eta_\pi 
(\eta_\pi/\eta)^{1/3}$ (Bahcall \& \Mesz 2000, Derishev et al 1999), while the
protons (which are the component capable of being heated by shocks) reach a maximum bulk 
Lorentz factor $\Gamma_{pf}\sim \eta_\ast < \eta$ (\Mesz \& Rees 2000). Shocks can still
occur, but the radius at which they occur reaches a maximum value $r_{sh,M}/r_o \sim (t_v/t_o) 
\eta_\ast^2 \simeq$ constant.

Below the radius $r_\pi$ the pionization optical depth $\tau'_\pi\simg 1$. The neutrino emission 
from inelastic $np$ collisions at threshold leading to pion and muon decay contains, on average, 
one $\num$ and one $\barnum$ per neutron, with mean CM energies of 30-50  MeV, and also either 
a $\nue$ or $\barnue$ with an average CM energy of about 30 MeV, resulting from pions and muons 
decaying at rest in the CM (Bahcall \& \Mesz, 2000). 
Typical conditions assumed for internal shocks involve collisions 
between shells with saturated Lorentz factors differing by factors of a few, e.g. a root-mean 
square value $\eta=\sqrt{\eta_A \eta_B}=10^2 \eta_2$ with $\eta_A=200,~\eta_B=50$ as an example.
If the luminosity is constant between variations, this corresponds to a mechanical
dissipation efficiency $\varep=(\eta_A+\eta_B-2\sqrt{\eta_A\eta_B})/(\eta_A+\eta_B)=0.2$.
The typical collision Lorentz factor is $\Gamma_{rel}\simeq (1/2)(\eta_A/\eta_B + 
\eta_B/\eta_A)\simeq 2.12$, and the total CM energy is $W=(2m_p^2 +2 m_p E_{rel})^{1/2}\sim 
2.35$ GeV, whereas the CM threshold energy is $2m_p+m_\pi\sim 2.01$ GeV. Thus, at depths
$\tau_\pi >1$ each $n$ heated by an individual shock can, on average, collide $k_\pi \sim 2$ 
times (and produce a pion each time) before its CM energy drops below threshold.
The observer-frame energy of the $\num\barnum$ ($\nue\barnue$) which would  be
predominant in the detector is $\sim 50(30) \alpha \eta/(1+z)$ MeV, where $\alpha\sim 1$  
near threshold, or
\beq
\eps_{\num\barnum} \sim  3~\alpha 
 \left({\eta \over \etapish}\right) \left({ 2 \over 1+z}\right)~\hbox{GeV}~~,~~
\eps_{\nue\barnue} \sim 2~\alpha
 \left({\eta \over \etapish}\right) \left({ 2 \over 1+z}\right)~\hbox{GeV}~~,~~
\label{eq:epsnush}
\enq
where $\etapish$ is given in equation (\ref{eq:etapish}).
The average total $\nu+\barnu$ CM energy per neutron at threshold is $\eps'_{\nu\barnu}\sim 
100 \alpha k_\pi$ MeV, which in the observer frame is $\eps_{\nu\barnu}\sim 0.1 \alpha k_\pi 
\eta (1+z)^{-1}$ GeV. The average neutrino detection cross section averaged over $\nu$ and 
$\barnu$ is (Gaisser, 1990) ${\sigma}_{\nu\barnu} \sim 0.5\times 10^{-38} (\eps_{\nu\barnu} /
\hbox{GeV})\hbox{cm}^2$, which (per neutron) is ${\bar \sigma}_{\nu\barnu} \sim 0.6\times 
10^{-37}\alpha (k_\pi /2) (\eta /\etapish) (2/1+z)$ cm$^2$.
The total number of neutrons involved in the shock and diffusion process, with a shock 
dissipation efficiency $\varep =0.2 \varep_{0.2}$, is 
\beq
N_{n,sh} = \left( {\xi \over 1+\xi} \right) \left( { \varep E \over \eta m_p c^2}\right)
 \sim 5.5 \times 10^{52} \varep_{0.2} E_{53} 
  \left({\eta \over \etapish}\right)^{-1} \left({2\xi \over 1+\xi }\right) ~.
\label{eq:Nnsh}
\enq
Assuming a km$^3$ detector with $N_t\sim 10^{39}N_{t39}$ target protons, and a burst rate out to a 
Hubble radius  of $10^3 {\cal R}_b$ per year, the number of events per year in the detector will 
be $R_{\nu\barnu}\sim (N_t/4\pi D^2){\cal R}_b N_n {\bar \sigma}_{\nu\barnu}$, where $D$ is the 
proper distance out to redshift $z$. For an Einstein-de Sitter universe with Hubble constant 
$H=65 h_{65}$ km/s/Mpc this is 
\beq
R_{\nu\barnu,sh} \sim 3~ E_{53} \varep_{0.2} \alpha \left({k_\pi \over 2}\right) 
  \left({2 \xi \over 1+\xi }\right) N_{t39} {\cal R}_{b3} 
  h_{65}^2 \left( {2- \sqrt{2} \over 1+z - \sqrt{1+z}}\right)^2~\hbox{year}^{-1}~,
\label{eq:Rnush}
\enq
in coincidence with observed GRBs, and independent of the value of $\eta$ for $\eta\siml
\eta_{\pi,sh}$, which is the limit below which $\tau_\pi \simg 1$. For values $\etapish
\siml \eta \siml \etapi$, the rate (\ref{eq:Rnush}) must be multiplied by $\tau'_\pi
\sim (\etapish/\eta)^5$, so between these limits $R_{\nu\barnu}$ drops steeply. However,
for $\eta\simg\etapi\sim 400$, even in a laminar flow without any shocks one expects radial 
decoupling of the $np$ leading to inelastic collisions, in which case the rate becomes a 
factor $(k_\pi\varep_{sh})^{-1} \sim 2.5$ larger, or (Bahcall \& \Mesz 2000) 
$R_{\nu\barnu}\sim 7$/year for typical $\eps_{\nu_\mu}\sim \eps_{\bar \nu_\mu} \sim 10$ GeV. 


\section{External neutron injection into the beam}
\label{sec:lateral}

If the progenitor is a collapsing massive star, a relativistic fireball from the collapsed 
central engine, probably in the form of a jet, must punch its way out through the stellar core 
and the envelope in order to become a `manifest' GRB which emits observable $\gamma$-rays. 
The collision of fast baryons from  the jet with quasistatic or slow neutrons  or baryons
from the rest of the star is a possible source of high energy neutrinos. We show here that
two of the most straightforward scenarios have a very low neutrino efficiency, and then go 
on to discuss a third  scenario where the efficiency is significantly higher.

The simplest scenario is that of a beam-dump provided by the outer envelope of a massive 
progenitor star, before the jet has punched through it. The fast neutrons of the jet would 
penetrate ahead of the protons participating in the bow shock, and collide inelastically 
with the upstream baryons at relative Lorentz factors $\sim \eta_j \sim 10^2-10^3$. However, a 
jet with such a Lorentz factor takes only $t_{cross} \sim r/c\eta^2\sim 3\times 10^{-2} r_{13}
\eta_2^{-2}$ s in the observer frame to reach an envelope radius $10^{13}r_{13}$ cm, and for 
a burst of observed duration $t_b =30(2/1+z)$ s only a fraction $(t_{cross}/t_b)\sim 2\times 
10^{-3} r_{13}\eta_2^{-2} (t_b/30)^{-1}$ of the jet neutrons will collide with envelope targets 
before the latter is punched through. Assuming that the observer is in the jet beam, the
isotropic equivalent output of jet neutrons is $N_n=3\times 10^{53}\E53 \eta_2^{-1}
(2\xi/1+\xi)$. The jet neutrons have a source frame energy $\eps_{j} \sim 10^2\eta_2$ GeV, 
well above the inelastic threshold for envelope baryons. Assuming that the pions decay 
at rest in the CM, the maximum observer frame energy of the $\num$ or $\barnum$ is 
$\eps_{\nu_\mu \bar \nu_\mu} \sim 50 \alpha \eta /(1+z)~\hbox{MeV} = 2.5 \alpha\eta_2(2/1+z)$ GeV. 
The average total $\nu\barnu$ energy per neutron is $\eps\sim 0.1\alpha\zeta_\pi \eta/(1+z)$ GeV 
and the total detection cross section per neutron is ${\bar \sigma}_{\nu\barnu} 
\sim 2.5\times 10^{-38} \alpha\zeta_\pi\eta_2(2/1+z)$ cm$^2$, where $\zeta_\pi$ is the pion 
multiplicity. The expected rate of events is  $ R_{\nu\barnu,dump} \sim (N_t/4\pi D^2){\cal R}_b 
{\bar \sigma}_{\nu\barnu} N_n (t_{cross}/t_b)$ $\sim 1.5\times 10^{-2} \E53 \alpha\zeta_\pi 
\eta_2^{-2} r_{13} (30\hbox{s}/t_b)(2/1+z)$ year$^{-1}$, with the same detector and cosmological 
scalings as in equation (\ref{eq:Rnush}). This is therefore too low for km$^3$ detectors.

A second natural scenario involves transverse neutron diffusion into the beam from the core of 
the evolved progenitor star, and illustrates some of the factors affecting neutrino efficiency, 
as well as serving as a preliminary to the third, more efficient scenario discussed in the 
subsequent paragraph. As the progenitor collapse proceeds the core will be heated by  
a shock moving outwards with $v= \beta_s c$ to temperatures characteristic of the virial
value near the central engine region, and it will be permeated by a diffuse bath of thermal 
neutrinos and gamma rays capable of dissociating the metals in the core. The free neutrons 
then diffuse via elastic nuclear collisions, into the jet channel. Taking a core neutron
density $n_n=\xi_c {\bar n}_c$ where ${\bar n}_c=3\times 10^{26} m_c r_{10}^{-3}$ cm$^{-3}$ is 
the baryon density for a core of $m_c$ solar masses, the neutron flux diffusing into the channel 
is $\phi_n \sim (v_t \ell /t)^{1/2} n_n$ cm$^{-2}$ s$^{-1}$ after a time $t$, where 
$\ell\sim ({\bar n}_c \sigma_{el})^{-1}$, and $\sigma_{el} \sim \sigma_\pi (c/v_t)$ is the 
elastic cross section, $v_t \sim \beta_s c$ being the neutron thermal velocity in the core. 
Using an effective transverse area of the channel of $A\sim \pi r^2 \theta_j$, the 
time-integrated fluence of neutrons into the channel after time $t$ is $N_{n,\perp} \sim 
2 t \pi r_c^2\theta_j \beta_s ( \xi / 1+\xi )(c {\bar n}_c /\sigma_\pi t)^{1/2}$, 
which is $N_{n,\perp} \sim 10^{51} \theta_{-1} r_{10}^{1/2} (m_c / 2)^{1/2} (\beta_s / 3) 
(2\xi / 1+\xi ) (t_b / 30\hbox{s})^{1/2} (2 / 1+z )^{1/2}$, using a core mass $m_c=2\msun$, 
an observed duration $t_b=30(2/1+z)$ s and a jet angle $\theta=10^{-1}\theta_{-1}$ radians. 
This is comparable to the number of baryons going out through the jet, which is $N_b \theta^2
\sim 6\times 10^{51}\E53\eta_2^{-1}\theta_{-1}^2$, where $N_b$ is the isotropic equivalent output 
of jet baryons, $N_b=6\times 10^{53}\E53 \eta_2^{-1}$.
The neutrons diffusing into the channel collide against the ultrarelativistic baryons in the jet, 
and as above, the maximum observed energy of the $\num$ or $\barnum$ produced is 
$\eps_{\nu_\mu \bar \nu_\mu} \sim 2.5 \alpha \eta_2(2/1+z)$ GeV and the average detection cross 
section per neutron is ${\bar \sigma}_{\nu\barnu} \sim 2.5\times 10^{-38} \alpha\zeta_\pi\eta_2
(2/1+z)$ cm$^2$.  However, the transverse inelastic optical depth through the jet encountered by 
the diffusing slow neutrons is $\tau_{\pi,\perp}\sim n_n\sigma_\pi r\theta \sim 5\times 10^4 \L52 
\eta_2^{-1} r_{10}^{-1} \theta_{-1}$. Thus only a fraction $\tau_{\pi,\perp}^{-1}$ of the jet 
baryons can produce 2.5 GeV neutrinos, since the slow neutrons cannot penetrate far into the jet.
The number of neutrino events in a km$^3$ detector is then $R_{\nu\barnu,core} \sim 
(N_t/4\pi D^2){\cal R}_b {\bar \sigma}_\nu 2 N_b \tau_{\pi,\perp}^{-1}$  
$\sim 1.5 \times 10^{-4} \eta_2 \alpha \zeta_\pi r_{10} \theta_{-1}^{-1} (2/1+z)^2$ year$^{-1}$, 
with detector and cosmological scalings as in equation (\ref{eq:Rnush}). This is far too low, 
and it indicates the critical role played by the transverse optical depth. 

A maximally efficient scenario based on lateral neutron injection can be obtained
when the jet transverse inelastic depth is of order unity, so that neutrons diffusing 
inwards can interact with the entire fast beam. The transverse inelastic depth of
the jet baryons in the lab frame is 
\beq
\tau_{\pi\perp}=n_j \sigma_\pi r\theta \sim \eta_\pi^4 \eta^{-1} \theta (\ro/r)~,
\label{eq:tauperp}
\enq
and unit optical depth is achieved at $r/\ro\sim 4\times 10^6\theta_{-1}\eta_3^{-1}\L52 r_{07}^{-1}$
or about $4\times 10^{13}$ cm for a jet with $\eta=10^3\eta_3 \sim 1$. If the jet is surrounded 
by a slower outflow, say with $\Gamma_s \sim 10$ (and $\Gamma_s \siml \theta_j^{-1}$), the 
slow outflow can reach a distance $4\times 10^{13}$ cm in an observer frame burst duration, 
$r/c\Gamma_s^2 \sim t_b \sim 30$ s, and internal shocks in the slow outflow can heat the baryons
in it. Inelastic collisions will heat the neutrons in the slow flow carried along from lower 
radii to thermal velocities $v_t\siml c$, and lead also to $\nu\barnu$, but their lab-frame 
energy will be very low, $\siml 0.25$ GeV. The lab frame density in the slow outflow will 
be $10^2$ times larger than in the jet (assuming the same energy per solid angle), and neutrons 
can diffuse into the jet from a depth in the slow outflow which is one-tenth of the width of the 
jet in the time taken for a neutron to cross the jet, the amount of slow mass in this slow width
being ten times that in the jet. If the pions and muons are produced at rest in the CM, using
the CM neutrino energies $\eps'$ of the previous section the maximum observer frame energy of the 
$\nu$ or $\barnum$ from inelastic collisions with jet baryons is $\eps\sim\eps' \eta(1+z)^{-1}$
\beq
\eps_{\num\barnum} \sim 25~\alpha \eta_3 \left({2 \over 1+z } \right) ~\hbox{GeV}~~,~~
\eps_{\nue\barnue} \sim 15~\alpha \eta_3 \left({2 \over 1+z } \right) ~\hbox{GeV}~,
\label{eq:epsnuj}
\enq
where $\alpha\sim 1$ near threshold.
The total observer frame neutrino energy per neutron is $\eps_\nu +\eps_{\bar \nu} \sim
0.1 \alpha \zeta_\pi \eta / (1+z)$ GeV, where $\zeta_\pi$ is the pion multiplicity, and the
average detection cross section averaged over $\nu$ and $\barnu$ is ${\bar \sigma}_{\nu\barnu}
\sim 2.5\times 10^{-37} \alpha \zeta_\pi \eta_3(2/1+z)$ cm$^2$ per neutron.
The (isotropic equivalent) baryon output in the jet is $N_b=E/\eta m_p c^2=6\times 10^{52}\E53
\eta_3^{-1}$, and the number of expected neutrino detection events is 
\beq
R_{\nu\barnu,j} \sim (N_t/4\pi D^2){\cal R}_b {\bar \sigma}_\nu 2 N_b 
 \sim 15 ~ \E53 \alpha \zeta_\pi N_{t39} {\cal R}_{b3} h_{65}^2 
 \left( {2- \sqrt{2} \over 1+z - \sqrt{1+z}}\right)^2~ \hbox{year}^{-1}~.
\label{eq:Rnuj}
\enq
This assumes that the maximum emission occurs at $\tau_{\pi\perp}\sim 1$, while for 
$\tau_{\pi\perp} \sim \E53 \eta_3^{-1} \theta_{-1} r_{13.6}^{-1}(t_b/30\hbox{s})^{-1}(1+z/2) >1$ 
the rate would be reduced by $\tau_{\pi\perp}^{-1}$. Note that if the transverse neutron input 
$N_{n\perp}\sim 2 \pi r^2 \theta \xi\beta_s (c t {\bar n}_s /\sigma_\pi)^{1/2}\sim 10^{53}\xi 
\beta_s\L52^{1/8} \eta_{s1}^{-1/2} t_1^{1/2}r_{07}^{3/8}r_{13.6}\theta_{-1}$ is less than the 
number of baryons in the jet $N_b\theta^2=6\times 10^{50}\E53\theta_{-1}^2\eta_3^{-1}$, then 
$N_{n\perp}$ must be used in equation (\ref{eq:Rnuj}), instead of $N_b$; this would be the case 
only if the wind neutron content is very small, $\xi/\theta_{-1} < 0.006 \L52^{7/8}t_1^{1/2}
\eta_3^{-1} \eta_{s1}^{1/2}\beta_s^{-1}r_{07}^{-3/8}r_{13.6}^{-1}$.

The above example used $\eta=10^3$, but the optimum efficiency $\tau_{\pi\perp} \simeq 1$ can
also be obtained for other values of $\eta$ in such a slow+fast jet model. The rate 
(\ref{eq:Rnuj}), $R_{\nu\barnu,j} \sim 15$/year, is independent of the value of $\eta$, since 
$\eta^{-1}$ in the number of ejected baryons cancels out with $\eta$ in the neutrino detection 
cross section. The neutrino energies do scale with $\eta$, and the same rate of 15/year would 
be obtained, e.g. for 
$\eta_j=10^2$ with $\eps_{\num\barnum} \sim 2.5$ GeV, $\eps_{\nue\barnue} \sim 1.5$ GeV, or for 
$\eta_j=10^4$ with $\eps_{\num\barnum} \sim 250$ GeV, $\eps_{\nue\barnue} \sim 150$ GeV.
The latter is an extreme value of $\eta_j$, but such energy neutrinos would be more easily
detectable with first generation km$^3$ detectors, which could therefore provide significant
constraints on the jet Lorentz factor.

\section{Discussion}

We have discussed two new mechanisms for producing significant fluences of 2-25 GeV neutrinos
in GRB fireballs. The first of these involves internal shocks occurring at or below 
the pionosphere (which itself is below the $\gamma$-ray photosphere), where the optical depth 
to inelastic nuclear collisions is large. This mechanism requires either temporal or angular
inhomogeneities (or both) in the relativistic outflow, which otherwise can be considered
as a simple flow (i.e. characterized by a single mean energy per solid angle and a single
mean Lorentz factor, aside from order unity variations in it). This mechanism can occur for 
values of $\eta \siml \eta_\pi^{4/5}(t_o/t_v)^{1/5}\siml 120 \L52^{1/5} r_{o7}^{-1/5}
(t_o/t_v)^{1/5}(2/[1+\xi])^{1/5}$. Thus, it considerably relaxes the $\eta$ parameter space 
requirements where neutrinos in the multi-GeV range can be expected.

The second mechanism involves transverse diffusion of neutrons from a slower outflow into a fast 
jet, at a height where the transverse inelastic optical depth of the latter is close to unity 
for maximal efficiency. This also requires the slow outflow to be heated by internal shocks
at or below its own pionosphere, and requires a two-zone outflow, characterized by two different
values of the mean Lorentz factor (or alternatively of the energy per solid angle).
This mechanism operates optimally at radii where the transverse optical depth is near unity, 
and predicts nominal event rates of $\siml 15$/year independent of $\eta$ and neutrino energies  
$\eps_{\num\barnum} \sim 25~ \eta_3 ({2/[1+z]} )$ GeV, $\eps_{\nue\barnue} \sim 15~ \eta_3 
({2 /[1+z]} )$ GeV which scale with $\eta$, providing a diagnostic for the value of the 
jet Lorentz factor.

A feature of these two mechanisms is that they can operate even if the outflow has initially 
a low neutron fraction, since inelastic collisions with optical depth $\simg 1$ tend to 
convert $p$ into $n$ and viceversa with approximately equal probability. They may therefore
occur either in neutron star mergers or in massive stellar progenitor collapses. 
For km$^3$ detectors such as ICECUBE (Halzen, 1999), ANTARES (Feinstein 1998), BAIKAL 
(Beloplatikov et al 1997) or NESTOR (Trascatti et al 1998) with sufficiently dense phototube 
spacing, the predicted event rates are $R_{\nu\barnu} \sim 3-15$/year, which would be detectable 
above the background in coincidence with GRB flashes. 
However, since these two mechanisms only require sub-photospheric internal shocks
(while longitudinal decoupling does not require any shocks), 2-25 GeV or higher energy
neutrinos could be observed from a much wider range of stellar collapse or compact merger 
events in the Universe at $z>1$, not necessarily associated with $\gamma$-ray flashes.

\acknowledgements{This research has been supported by NASA NAG5-2857, the Guggenheim
Foundation, the Sackler Foundation and the Royal Society. We are grateful to John Bahcall
for valuable comments and discussions.}

\begin{figure}[htb]
\centering
\epsfig{figure=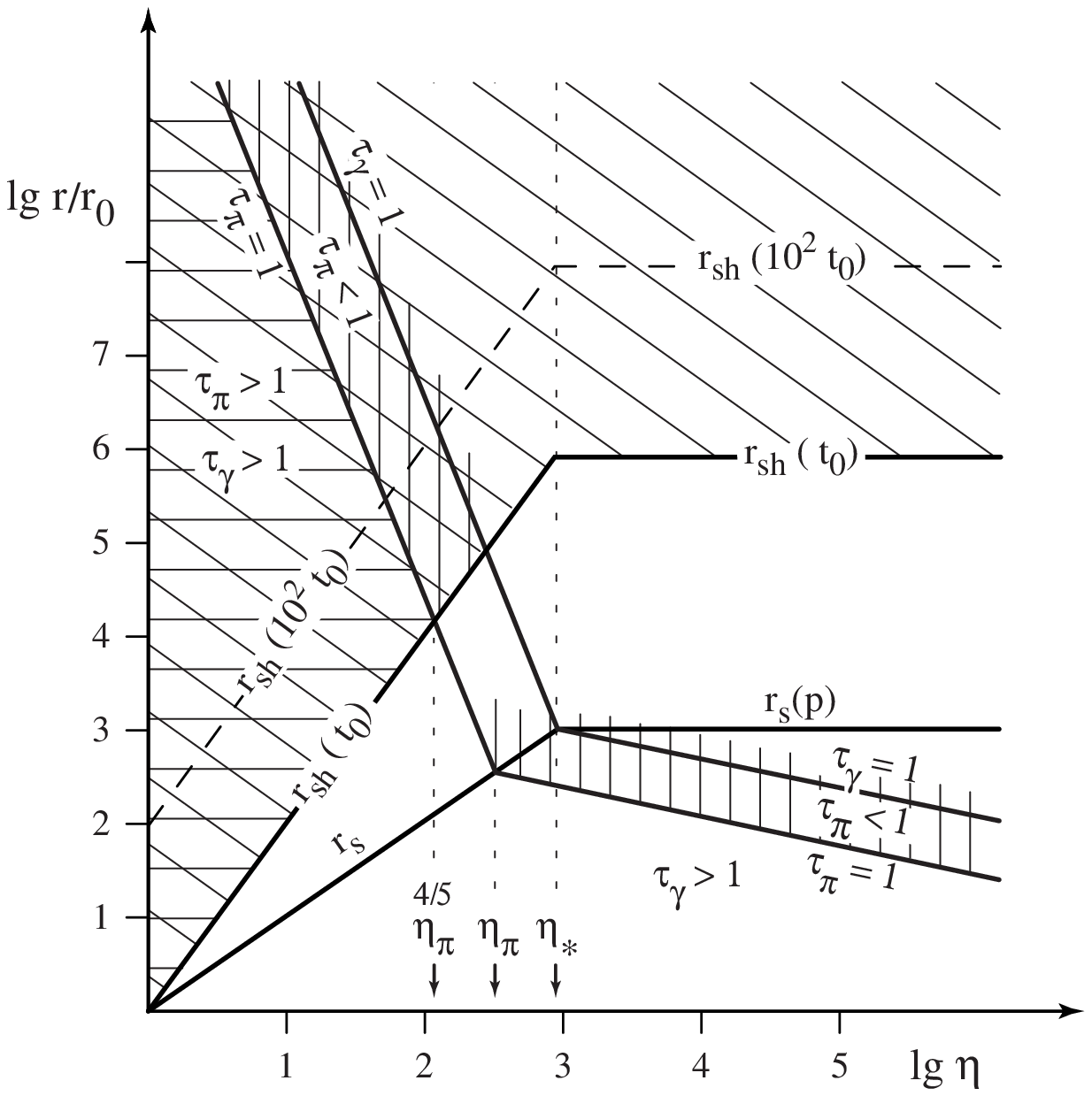, width=5.in, height=6.in}
\caption{ Radial distance vs. bulk Lorentz factor $\eta$ parameter space,
showing the pionosphere (pionization optical depth $\tau_\pi =1$),
and the photosphere (scattering optical depth $\tau_\gamma = 1$).
Horizontal hashed regions indicate $\tau_\pi>1$, this being absent above 
$\eta_\pi$ (where $n$ and $p$ are elastically coupled, until decoupling and
pionization occur at $\tau_\pi \simg 1$). Vertical hashed regions, only shown 
for part of the range, indicate $\tau_\pi <1$. The bulk Lorentz factor
saturation radius $r_s$ is common for $n$ and $p$ below $\eta_\pi$, while
above $\eta_\pi$ protons saturate at $r_s(p)$ and neutrons saturate (decouple)
at $\tau_\pi=1$. Internal shocks are possible above the line $r_{sh}(t_o)$,
marked by diagonal hashes. }
   \label{fig:tau}
\end{figure}

\end{document}